\documentclass[runningheads]{llncs}
\usepackage[T1]{fontenc}
\usepackage{graphicx}
\usepackage[numbers,sort&compress]{natbib}
\usepackage{algorithm}
\usepackage[noend]{algpseudocode}
\usepackage{siunitx}
\usepackage{url}
\usepackage{pifont}
\usepackage{xcolor}
\usepackage{comment}
\usepackage{multirow, makecell}
\usepackage{enumitem}
\usepackage{amsfonts}
\usepackage{booktabs}
\usepackage{needspace}
\usepackage{amsmath}
\usepackage{orcidlink}
\begin{document}
\title{LibriConvo: Simulating Conversations from Read Literature for ASR and Diarization}
\titlerunning{LibriConvo}
%
\author{Máté Gedeon\inst{1,2}\orcidlink{0009-0005-1429-8279} \and Péter Mihajlik\inst{2,3}\orcidlink{0000-0001-7532-9773}}
\authorrunning{M. Gedeon, P. Mihajlik}
%
\institute{Department of Telecommunications and Artificial Intelligence, \\ Budapest University of Technology and Economics, Hungary \and
Speechtex Ltd., Hungary \and
ELTE Research Centre for Linguistics, Hungary \\
\email{gedeonm@edu.bme.hu, mihajlik@tmit.bme.hu}}

\maketitle              
\begin{abstract}
We introduce LibriConvo, a synthetic conversational speech corpus for speaker diarization and automatic speech recognition (ASR), built by instantiating the previously proposed Speaker-Aware Simulated Conversation (SASC) framework in a dataset and benchmarking setting. The main contribution of this paper is a corpus construction pipeline and benchmark derived from that framework. To make the data more suitable for downstream ASR and diarization, conversational timing statistics are estimated from English CallHome using external voice activity detection, long pauses are compressed, LibriTTS utterances are grouped by book to improve local semantic continuity, and room impulse responses are selected with a spatial-plausibility heuristic. The resulting corpus contains 240.1 hours of audio across 1,496 dialogues involving 830 speakers, partitioned into speaker-disjoint train, validation, and test splits. We report baseline results for both diarization and ASR. On the test split, Sortformer outperforms the pyannote pipeline in diarization (11.1\% vs.~24.4\% DER). For ASR, a Fast Conformer-CTC XLarge model fine-tuned with Serialized Output Training achieves 7.29\% WER and 6.97\% cpWER, outperforming zero-shot Whisper-large-v3. These results position LibriConvo as a practical benchmark for studying synthetic conversational speech and for evaluating multi-speaker speech processing systems.

\end{abstract}

\section{Introduction}
Modern speech processing systems, including end-to-end speaker diarization (EEND) and multi-speaker automatic speech recognition (ASR), require large amounts of annotated conversational audio \cite{CHiME-6}. Collecting real multi-party conversations with reliable turn-level labels is expensive and difficult, so synthetic data generation has become a common alternative \cite{Fujita2019}. Simulated mixtures provide exact supervision (\textit{who speaks when and what}) and are widely used for training diarization and ASR models \cite{Yu2016PIT, SOT}. However, naive utterance mixing often fails to capture realistic turn-taking behavior and stable speaker identity patterns across a full dialogue \cite{Landini2022MultiSpeakerEEND}.

Recent work has moved toward more naturalistic simulation. \citet{Yamashita2022Naturalness} explicitly model transition types to better match silence/overlap statistics. \citet{Landini2022} create conversation-like mixtures using pause and overlap distributions estimated from real dialogue. \citet{Park2023} propose a probabilistic \emph{property-aware} simulator to control target interaction statistics.

A recent study introduced the Speaker-Aware Simulated Conversation (SASC) framework \cite{SASC}, which models conversational timing with unified gap/overlap distributions, speaker-dependent temporal variation, and Markov-chain turn-taking. It contributed the simulation framework and intrinsic analysis, but did not provide downstream ASR and diarization experiment results, nor a released benchmark dataset.

This paper focuses on that missing step: turning the SASC idea into a usable corpus. The core simulation principle comes from prior work \cite{SASC}; the contribution here is the dataset-oriented instantiation and evaluation. Specifically, this paper introduces LibriConvo, a 240.1-hour corpus (1,496 dialogues, 830 speakers) with speaker-disjoint train/validation/test splits, designed for both diarization and ASR training enhancement and benchmarking.

The main contributions of this paper are as follows:
\begin{itemize}
\item We introduce \textbf{LibriConvo}, a synthetic conversational speech dataset for both ASR and speaker diarization, with turn-level speaker labels and transcriptions.
\item We provide a \textbf{dataset-construction pipeline} based on SASC, including (i) book-constrained utterance sampling to improve semantic continuity and (ii) a realism-driven room impulse response (RIR) selection strategy for more plausible acoustics.
\item We establish \textbf{reproducible baselines} for both tasks by reporting diarization and ASR results on the speaker-disjoint splits.
\end{itemize}
The dataset will be released upon publication.

The structure of the paper is as follows. Section 2 introduces the theoretical framework of the proposed methodology and its application to dataset generation. Section 3 presents the baseline diarization and ASR results. Finally, Section 4 concludes the paper by summarizing the findings and outlining directions for future work.

\section{Methodology}
\subsection{Speaker-aware conversation simulation}

In this subsection, we summarize the speaker-aware conversation simulation (SASC) method introduced by \citet{SASC},which generates multi-speaker dialogues with temporal, structural, and acoustic properties that are modeled based on real conversations. Conversational timing is represented by a unified distribution of gaps $\delta$, where $\delta < 0$ indicates overlap, $\delta \geq 0$ indicates a pause, and the integral over the negative domain corresponds to the probability of overlap $p_{\text{overlap}}$. Instead of parametric or histogram-based approaches, kernel density estimation (KDE) is used to obtain smooth, continuous estimates of these gap distributions.

For timing consistency, two mean pause distributions are defined: $\hat{D}_{=}$ for same-speaker mean gaps (when no speaker change occurs between utterances) and $\hat{D}_{\neq}$ for different-speaker mean gaps. For each speaker $s$, an initial base value is sampled from the appropriate distribution, while subsequent gaps are generated by adding a deviation:
\[
\delta_n =
\begin{cases}
\mu^{\text{same}}_s + v & \text{if } X_n = X_{n-1},\; v \sim V_{=} ,\\[4pt]
\mu^{\text{diff}}_s + v & \text{if } X_n \neq X_{n-1},\; v \sim V_{\neq}.
\end{cases}
\]

Here $V_{=}$ and $V_{\neq}$ are zero-mean speaker deviation distributions that preserve local temporal consistency across turns.

Turn-taking is modeled by a first-order (generalizable to $n$-order) Markov chain with transition matrix $P_{\mathrm{turn}}$, which defines the probability of selecting the next speaker given the previous one. All speakers are placed within a single acoustic environment by sampling a room from the available RIRs and assigning distinct positions within that room. After all utterances are concatenated with their respective gaps and overlaps, optionally background noise $n \sim \mathcal{N}$ is added and scaled according to a sampled signal-to-noise ratio $r \sim \mathcal{R}$ to produce the final mixture. A compressed version of the procedure is shown in Algorithm~\ref{sasc_algo}.

\begin{algorithm}[h]
\label{sasc_algo}
\caption{Simplified speaker-aware conversation simulation}
\begin{algorithmic}[1]
\State Select $N_{\mathrm{spk}}$ speakers $\mathcal{S}'$; assign RIRs (same room, distinct positions)
\State Choose initial speaker $X_1$
\For{$n = 1 \dots N_{\mathrm{u}}$}
  \If{$n>1$} sample $X_n \sim P_{\mathrm{turn}}(X_{n-1})$ \EndIf
  \State Sample utterance $u_n \in U_{X_n}$, convolve with RIR $\to y_n$
  \If{$n=1$} set $\delta=0$
  \ElsIf{$X_n = X_{n-1}$}
     \If{first gap for $X_n$} $\mu^{\text{same}}_s \sim \hat{D}_{=}$ \EndIf
     \State $\delta = \mu^{\text{same}}_s + v,\; v \sim V_{=}$
  \Else
     \If{first gap for $X_n$} $\mu^{\text{diff}}_s \sim \hat{D}_{\neq}$ \EndIf
     \State $\delta = \mu^{\text{diff}}_s + v,\; v \sim V_{\neq}$
  \EndIf
  \State Mix $y_n$ into conversation with gap $\delta$
\EndFor
\end{algorithmic}
\end{algorithm}

\subsection{Dataset generation}
To move beyond the theoretical framework, and implement the methodology, we first define the datasets and methods employed.

\subsubsection{Statistics}
In their original work, \citet{SASC} noted that the annotations for Switchboard \cite{switchboard} exhibited inconsistencies, reporting an average gap corresponding to approximately half a second of overlap—an outcome that appears implausible in natural conversational settings. To mitigate this issue, we opted to use the English CallHome corpus \cite{CallHome}, complemented by an external voice activity detection (VAD) model -- Silero VAD~\cite{Silero} -- to obtain more reliable temporal boundaries.

\begin{figure}[h]
    \centering
    \includegraphics[width=0.7\linewidth]{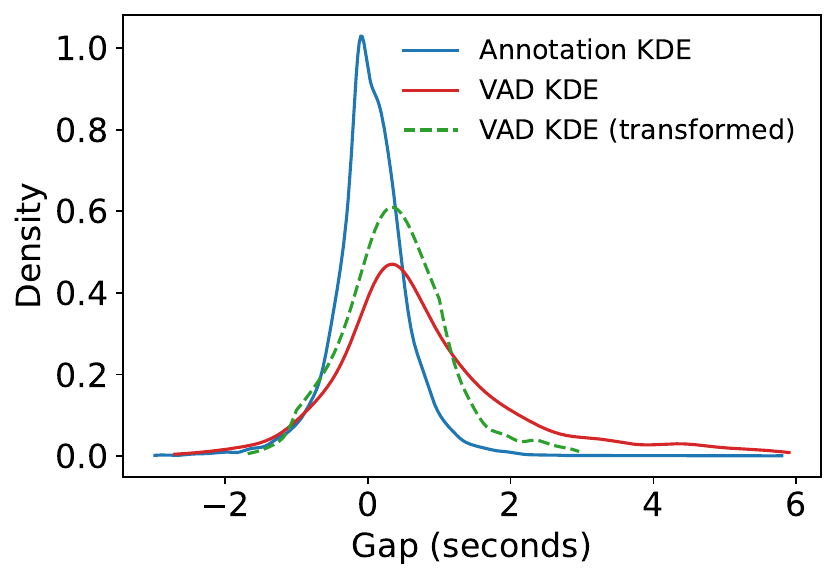}
    \caption{Comparison between original CallHome annotations and VAD-derived speech segments.}
    \label{fig:vad_plot}
\end{figure}

Even with CallHome, notable discrepancies emerged between the original annotations\footnote{\url{https://huggingface.co/datasets/talkbank/callhome}} and the VAD-derived speech segments, as illustrated in Figure~\ref{fig:vad_plot}. This contrast likely stems from the fact that the CallHome annotations were never intended for precise gap or overlap analysis, but rather to provide approximate timestamps suitable for diarization training.

Preliminary listening to reconstructed conversations revealed that the dialogues felt disjointed, with pauses longer than typical in spontaneous speech. To better emulate natural conversational timing, we made the subjective choice to apply a temporal compression to the detected pauses, which selectively shortens longer silences while preserving short gaps. More specifically, we applied a deterministic piecewise linear pause-compression transform to the detected silence durations. Let \(x \in \mathbb{R}\) denote a pause value (in standardized time units), \(t>0\) a threshold defining the range of short pauses to preserve, and \(s \in (0,1)\) a compression coefficient controlling the reduction of longer gaps. The transformed pause duration \(f(x)\) is defined as

\[
f(x)=
\begin{cases}
x, & |x|\le t \\
t+s(x-t), & x>t \\
-t+s(x+t), & x<-t
\end{cases}
\]

where in our implementation we used \(t=1.0\) and \(s=0.4\). Thus, pauses whose magnitude falls inside the interval \([-t,t]\) remain unchanged, while values outside this region are linearly compressed toward the threshold boundary with slope \(s\). Because \(s<1\), increasingly long pauses are proportionally shortened, yet their ordering and relative distances are preserved.

For turn-taking modeling, we used a Markov chain of order 1, based on the CallHome data.

\subsubsection{Utterances}
A common limitation in prior studies is that little attention was given to the semantic consistency of the texts used for simulated conversations. While this poses only a minor concern for EEND, it can significantly hinder ASR models, which can benefit from textual context to refine recognition. However, sourcing independent spoken texts that are meaningfully aligned is challenging. To make sure utterances have some semantic common ground, we used utterances that are read parts of a book, with a fixed book for each simulated conversation. The utterances came from LibriTTS \cite{LibriTTS}, which is already segmented to sentences. To make utterance length realistic in the conversations, we used utterances from 2 seconds to 10 seconds of length.

\subsubsection{Room impulse responses}
For the room impulse responses (RIRs), we used the BUT Speech@FIT Reverb Database \cite{BUTdb}, which provides recordings from nine rooms with 31 microphones and about five speakers per room—an ideal configuration for our use case. A key consideration, however, is to select microphones that reflect a realistic acoustic setup—specifically, avoiding positions mounted on ceilings or walls, which would not typically occur in practical scenarios.

To achieve this, we designed a RIR selection procedure that ranks all speaker–microphone configurations within each room by their spatial plausibility. Each configuration is described by four geometric attributes—height, distance, elevation, and azimuth—relative to the microphone array. A realism score is then computed as a weighted sum of normalized deviations from idealized reference values: 1.5 m speaker (source) height, 1 m source–microphone distance, and 0° elevation. Lower scores indicate configurations closer to typical human speech positions.

To maintain spatial diversity, we selected multiple speakers with azimuths differing by at least 20°, avoiding collinearity and better approximating conversational scenes. For each selected speaker, an RIR was randomly drawn from its associated microphone positions. This realism- and diversity-driven strategy yields acoustically plausible and varied RIRs consistent with natural recording conditions.

Overall, the method acts as a lightweight spatial optimization that embeds perceptually motivated heuristics into the RIR sampling process. Unlike prior work optimizing placement for intelligibility or coverage \citep{Morales2019}, our approach targets achieving data realism after the recordings happened, automatically filtering implausible configurations without manual inspection or exhaustive simulation. We applied this procedure to 40\% of the conversations.

\subsubsection{Creating splits}
To ensure robust evaluation and eliminate any speaker overlap between training and testing, we constructed speaker-disjoint splits of the simulated conversations into training, validation, and test sets. Unique speakers were first extracted from the metadata, and all their associated conversations were identified. Since each conversation involved two participants, it was insufficient to simply distribute individual speakers into subsets; speaker pairings also had to be taken into account. Consequently, speakers were heuristically grouped and randomly assigned to one of three mutually exclusive subsets. Conversations were then allocated based on speaker membership to approximate an 80–10–10\% train–validation–test ratio. This procedure preserves conversational integrity, maintains balanced data proportions, and ensures that no speaker identity appears in more than one subset, enabling fair generalization to unseen speakers. The resulting distribution is presented in Table~\ref{tab:split_stats}.

\begin{table}[h]
\centering
\resizebox{0.7\columnwidth}{!}{
\begin{tabular}{lcccc}
\toprule
\textbf{Subset} & \textbf{Speakers} & \textbf{Conversations} & \textbf{Duration (h)} \\
\midrule
Train & 580 & 1199 & 193.7 \\
Validation & 127 & 137 & 23.1 \\
Test & 123 & 160 & 23.3 \\
\midrule
\textbf{Total} & 830 & 1,496 & 240.1 \\
\bottomrule
\end{tabular}}
\vspace{5pt}
\caption{Dataset split statistics.}
\label{tab:split_stats}
\end{table}

\section{Experiments}
We conducted evaluations on the generated dataset, providing useful baselines for further research.

\subsection{Data preparation}
To facilitate efficient processing and model training, simulated conversations were segmented into temporally coherent units of up to 30 seconds, matching the input limit of the \texttt{Whisper-large-v3\footnote{\url{https://huggingface.co/openai/whisper-large-v3}}}
 ASR model. Utterances were added sequentially until the next would exceed this limit, at which point a new segment was initiated, omitting intervening silences. Segment times were redefined relative to onset while preserving absolute timestamps. This procedure maintains natural temporal continuity and provides consistent analysis units for training, evaluation, and error tracking.

\subsection{Diarization}
We evaluated speaker diarization performance on our validation and test sets using two state-of-the-art (SOTA) diarization frameworks, both applied without additional fine-tuning to assess their generalization capabilities on our dataset. The goal of this evaluation was to establish baseline results and analyze how different model architectures perform, when confronted with simulated multi-speaker speech.

The first baseline is the \textit{pyannote} diarization pipeline \cite{pyannote}, a modular framework composed of neural components for speech segmentation, speaker embedding extraction, and clustering. We employed the \texttt{speaker-diarization-3.1}\footnote{\url{https://huggingface.co/pyannote/speaker-diarization-3.1}}
 model. The system detects short speech segments using a sliding-window segmentation module, extracts speaker-discriminative embeddings from each, and groups them via agglomerative hierarchical clustering. Finally, clustered segments are merged in a post-processing step to produce the diarization output.

The second baseline is \textit{Sortformer} \cite{Sortformer}, an encoder-based diarization model originally designed to supervise speaker tagging in speech-to-text systems. Unlike traditional methods relying on permutation-invariant loss, Sortformer introduces a \emph{Sort Loss} that enforces a consistent ordering of speaker labels, jointly modeling speaker assignment and temporal continuity. This design improves both diarization accuracy and multi-speaker transcription by embedding speaker identity information directly into the ASR process. We used the \texttt{diar\_sortformer\_4spk-v1}\footnote{\url{https://huggingface.co/nvidia/diar_sortformer_4spk-v1}} model.

Table~\ref{tab:diar_comparison} presents the average diarization error rate (DER) obtained with both models on the evaluation and test sets. As shown, Sortformer significantly outperforms the pyannote pipeline, achieving a considerably lower DER on both sets. This suggests that the end-to-end transformer architecture is more effective at disentangling overlapping speech and maintaining speaker consistency over longer segments.

\begin{table}[ht]
\centering
\resizebox{0.6\columnwidth}{!}{
\begin{tabular}{lcc}
\toprule
\textbf{Model} & \textbf{Validation (\%)} & \textbf{Test (\%)} \\
\midrule
Pyannote      & 25.6 & 24.4 \\
Sortformer    & \textbf{12.9} & \textbf{11.1} \\
\bottomrule
\end{tabular}}
\vspace{5pt}
\caption{Comparison of diarization models.}
\label{tab:diar_comparison}
\end{table}

\begin{figure}[h]
    \centering
    \includegraphics[width=0.7\linewidth]{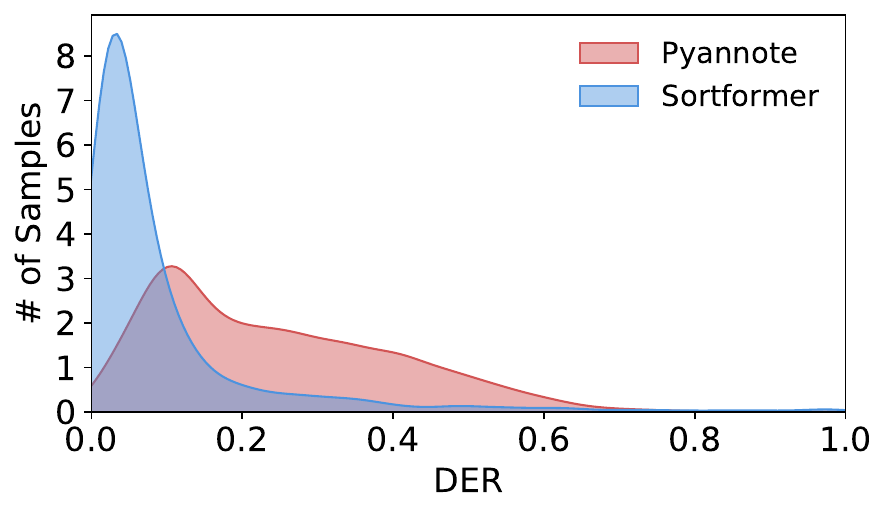}
    \caption{Distribution of DER values across test recordings.}
    \label{fig:der_density}
\end{figure}

To further analyze performance variability, Figure~\ref{fig:der_density} visualizes the distribution of DER values across individual recordings in the test set. In addition to achieving a substantially lower mean error, Sortformer demonstrates much higher consistency, as indicated by the narrower spread of its error density. In contrast, the pyannote pipeline shows larger variance, often struggling with recordings containing high speaker overlap or rapid turn-taking.

\begin{table*}[t!]
\centering
\label{tab:asr_results}
\resizebox{1.0\columnwidth}{!}{
\begin{tabular}{lcccccc}
\toprule
\multirow{2}{*}{\textbf{Model}} & 
\multicolumn{3}{c}{\textbf{Validation}} & 
\multicolumn{3}{c}{\textbf{Test}} \\
\cmidrule(lr){2-4} \cmidrule(lr){5-7}
& \textbf{WER $\downarrow$} & \textbf{cpWER $\downarrow$} & \textbf{SegAcc $\uparrow$}
& \textbf{WER $\downarrow$} & \textbf{cpWER $\downarrow$} & \textbf{SegAcc $\uparrow$} \\
\midrule
\texttt{Whisper-large-v3} & 9.41 & 9.25 & - & 7.46 & 7.30 & - \\
\texttt{Canary-1b-v2} & \textbf{8.69} & 8.69 & - & 7.59 & 7.52 & - \\
\texttt{fastconformer\_l} & 22.57 & 22.41 & - & 23.14 & 23.07 & - \\

\texttt{fastconformer\_xl} & 16.98 & 16.87 & - & 16.82 & 16.76 & - \\
\midrule
\texttt{fastconformer\_l (ft)} & 10.08 & 9.89 & 84.30 & 10.34 & 10.01 & \textbf{82.70} \\
\texttt{fastconformer\_xl (ft)} & 8.88 & \textbf{8.67} & \textbf{85.11} & \textbf{7.29} & \textbf{6.97} & 82.38 \\
\bottomrule
\end{tabular}}
\caption{ASR baseline results (\%) on validation and test splits, without and with fine-tuning.}
\end{table*}
\subsection{ASR}

To establish ASR baselines, we evaluated both state-of-the-art (SOTA) pretrained models—\texttt{Whisper-large-v3}~\cite{whisper} and \texttt{Canary-1B-v2}~\cite{canary-v2}—without fine-tuning, as well as smaller architectures based on \textit{FastConformer}~\cite{fastconformer} that were fine-tuned for our specific task. Training was performed using \textit{Serialized Output Training} (SOT)~\cite{SOT}, where speaker changes are explicitly marked with a \texttt{<sc>} (speaker change) token.

Unlike the token-level variant (\textit{t-SOT}) \cite{t-sot}, which signals every speaker change immediately on the token level with a \texttt{<cc>} (channel change) token, our approach treats overlapping speech differently. Specifically, we preserve each speaker’s utterance as a coherent unit, without splitting it when another speaker begins to overlap. 

To illustrate the conceptual difference between t-SOT and our SOT approach, consider the following example, where the word "I'm" starts before "you" and "good" starts before "thanks":

\begin{quote}
\textbf{Original:} \textit{-- How are you? -- I’m fine, thanks -- good}\\[4pt]
\textbf{t-SOT:} \texttt{How are <cc> I'm <cc> you? <cc> fine <cc> good <sc> thanks}\\[4pt]
\textbf{Ours:} \texttt{How are you? <sc> I'm fine, thanks <sc> good}
\end{quote}

For evaluation, we computed both the conventional \textit{Word Error Rate} (WER) and the \textit{concatenated minimum-permutation WER} (cpWER), which accounts for all possible permutations of segments separated by \texttt{<sc>} tokens and reports the minimal achievable WER. In addition, for the fine-tuned models, we measured \textit{segment accuracy (SegAcc)}, defined as the percentage of cases where the model correctly predicted the number of segments separated by \texttt{<sc>} tokens. Table~\ref{tab:asr_results} summarizes the evaluation metrics and outcomes. The two models used for fine-tuning were \texttt{fastconformer-ctc-large}\footnote{\url{https://huggingface.co/nvidia/stt_en_fastconformer_ctc_large}} and \texttt{fastconformer-ctc-xlarge}\footnote{\url{https://huggingface.co/nvidia/stt_en_fastconformer_ctc_xlarge}}.

As shown in Table~\ref{tab:asr_results}, the SOTA pretrained models achieve strong performance without fine-tuning. However, after task-specific fine-tuning, the \texttt{fastconformer\_ctc\_xlarge} model achieves the best overall results. Notably, all models show improvement when evaluated with the cpWER metric, suggesting that while their transcriptions are accurate in content, sometimes the ordering of speaker segments may differ from the reference annotations.

\section{Conclusion}
This paper presented the implementation of the previously proposed SASC framework as a conversational-speech corpus and benchmark for speaker diarization and automatic speech recognition (ASR), and evaluates the resulting dataset using strong baseline systems. It addressed key limitations of earlier simulated multi-speaker conversational datasets, particularly the lack of semantic coherence in utterance selection and the presence of implausible temporal gaps in annotated data. To overcome these issues, we developed a pipeline designed to generate conversational simulations that better support the training and evaluation of both diarization and ASR systems.

Our approach leverages the CallHome corpus with external VAD-based temporal boundary detection, applies temporal compression to reduce unnaturally long silences, and incorporates semantically coherent utterances from LibriTTS organized by source text. The integration of physically plausible room impulse responses through a realism-based selection strategy further enhances the acoustic authenticity of the generated data. The resulting dataset -- named LibriConvo -- comprises 240.1 hours of simulated conversations across 1,496 dialogues with 830 unique speakers, organized into speaker-disjoint train, validation, and test splits.

Baseline evaluations demonstrate the dataset's utility for benchmarking state-of-the-art systems. For diarization, Sortformer substantially outperformed the pyannote pipeline, achieving both lower mean DER and greater consistency across recordings, highlighting the advantages of end-to-end transformer architectures for handling overlapping speech and speaker continuity. For ASR, our Serialized Output Training approach—which preserves utterance-level coherence rather than fragmenting at every overlap—proved effective when combined with fine-tuning. The \texttt{fastconformer\_ctc\_xlarge} model achieved the best overall performance with 6.97\% cpWER on the test set after fine-tuning, outperforming SOTA pretrained models in our evaluation framework.

The strong baseline results and the dataset's realistic conversational characteristics position it as a valuable resource for future research in multi-speaker speech processing. The methodology presented here is reproducible and can be extended to generate larger-scale datasets or adapted to incorporate additional acoustic conditions and conversational patterns. Future work will explore the utility of datasets generated this way, when mixing with authentic conversational training data.

\begin{credits}
\subsubsection{\ackname} Project No. 2025-2.1.2-EKÖP-KDP-2025-00005 has been implemented with the support provided by the Ministry of Culture and Innovation of Hungary from the National Research, Development and Innovation Fund, financed under the EKÖP\_KDP-25-1-BME-21 funding scheme.
\end{credits}
%
%
%
\bibliographystyle{plainnat}
\bibliography{main}

@misc{CallHome,
  author       = {Canavan, Alexandra and Graff, David and Zipperlen, George},
  title        = {CALLHOME American English Speech},
  howpublished = {Web Download},
  address      = {Philadelphia},
  publisher    = {Linguistic Data Consortium},
  year         = {1997},
  note         = {LDC Catalog No.: LDC97S42, ISBN: 1-58563-111-6, ISLRN: 952-976-147-406-5},
  doi          = {10.35111/exq3-x930},
  url          = {https://catalog.ldc.upenn.edu/LDC97S42}
}

@article{BUTdb,
author = {Szoke, Igor and Skacel, Miroslav and Mosner, Ladislav and Paliesek, Jakub and Cernocky, Jan},
year = {2019},
month = {05},
pages = {1-1},
title = {Building and Evaluation of a Real Room Impulse Response Dataset},
volume = {PP},
journal = {IEEE Journal of Selected Topics in Signal Processing},
doi = {10.1109/JSTSP.2019.2917582}
}

@inproceedings{LibriTTS,
  title     = {LibriTTS: A Corpus Derived from LibriSpeech for Text-to-Speech},
  author    = {Heiga Zen and Viet Dang and Rob Clark and Yu Zhang and Ron J. Weiss and Ye Jia and Zhifeng Chen and Yonghui Wu},
  year      = {2019},
  booktitle = {Interspeech 2019},
  pages     = {1526--1530},
  doi       = {10.21437/Interspeech.2019-2441},
  issn      = {2958-1796},
}

@inproceedings{switchboard,
author = {Godfrey, John J. and Holliman, Edward C. and McDaniel, Jane},
title = {SWITCHBOARD: telephone speech corpus for research and development},
year = {1992},
isbn = {0780305329},
publisher = {IEEE Computer Society},
address = {USA},
booktitle = {Proceedings of the 1992 IEEE International Conference on Acoustics, Speech and Signal Processing - Volume 1},
pages = {517–520},
numpages = {4},
location = {San Francisco, California},
series = {ICASSP'92}
}

@inproceedings{CHiME-6,
  title     = {CHiME-6 Challenge: Tackling Multispeaker Speech Recognition for Unsegmented Recordings},
  author    = {Shinji Watanabe et al.},
  year      = {2020},
  booktitle = {6th International Workshop on Speech Processing in Everyday Environments (CHiME 2020)},
  pages     = {1--7},
  doi       = {10.21437/CHiME.2020-1},
}

@inproceedings{Park2023,
  title={Property-Aware Multi-Speaker Data Simulation: A Probabilistic Modelling Technique for Synthetic Data Generation},
  author={Park, T.J. and Huang, H. and Hooper, C. and Koluguri, N.R. and Dhawan, K. and Juki{\'c}, A. and Balam, J. and Ginsburg, B.},
  booktitle={Proc. 7th International Workshop on Speech Processing in Everyday Environments (CHiME 2023)},
  pages={82--86},
  year={2023},
  doi={10.21437/CHiME.2023-16}
}

@inproceedings{Landini2022,
  title={From Simulated Mixtures to Simulated Conversations as Training Data for End-to-End Neural Diarization},
  author={Federico Landini and Alicia Lozano-Diez and Mireia D{\'i}ez and Luk{\'a}š Burget},
  booktitle={Interspeech},
  year={2022},
  url={https://api.semanticscholar.org/CorpusID:247939646}
}

@article{Landini2022MultiSpeakerEEND,
  title={Multi-Speaker and Wide-Band Simulated Conversations as Training Data for End-to-End Neural Diarization},
  author={Federico Landini and Mireia D{\'i}ez and Alicia Lozano-Diez and Luk{\'a}š Burget},
  journal={ICASSP 2023},
  year={2022},
  pages={1-5},
  url={https://api.semanticscholar.org/CorpusID:253510723}
}

@inproceedings{Yamashita2022Naturalness,
  title={Improving the Naturalness of Simulated Conversations for End-to-End Neural Diarization},
  author={Natsuo Yamashita and Shota Horiguchi and Takeshi Homma},
  booktitle={The Speaker and Language Recognition Workshop},
  year={2022},
  url={https://api.semanticscholar.org/CorpusID:248377558}
}

@inproceedings{Fujita2019,
  title={End-to-End Neural Speaker Diarization with Permutation-Free Objectives},
  author={Yusuke Fujita and Naoyuki Kanda and Shota Horiguchi and Kenji Nagamatsu and Shinji Watanabe},
  booktitle={Interspeech},
  year={2019},
  url={https://api.semanticscholar.org/CorpusID:202572807}
}

@inproceedings{SOT,
  title={Serialized Output Training for End-to-End Overlapped Speech Recognition},
  author={Naoyuki Kanda and Yashesh Gaur and Xiaofei Wang and Zhong Meng and Takuya Yoshioka},
  booktitle={Interspeech},
  year={2020},
  url={https://api.semanticscholar.org/CorpusID:214714409}
}

@article{Yu2016PIT,
  title={Permutation invariant training of deep models for speaker-independent multi-talker speech separation},
  author={Dong Yu and Morten Kolb{\ae}k and Zheng-Hua Tan and Jesper H{\o}jvang Jensen},
  journal={ICASSP 2017},
  year={2016},
  pages={241-245},
  url={https://api.semanticscholar.org/CorpusID:7331600}
}

@misc{SASC,
      title={From Independence to Interaction: Speaker-Aware Simulation of Multi-Speaker Conversational Timing}, 
      author={Máté Gedeon and Péter Mihajlik},
      year={2025},
      eprint={2509.15808},
      archivePrefix={arXiv},
      primaryClass={cs.SD},
      url={https://arxiv.org/abs/2509.15808}, 
}

@article{Morales2019,
title = {Receiver placement for speech enhancement using sound propagation optimization},
journal = {Applied Acoustics},
volume = {155},
pages = {53-62},
year = {2019},
issn = {0003-682X},
doi = {https://doi.org/10.1016/j.apacoust.2019.04.037},
url = {https://www.sciencedirect.com/science/article/pii/S0003682X18303657},
author = {Nicolas Morales and Zhenyu Tang and Dinesh Manocha}
}

@inproceedings{pyannote,
  title     = {pyannote.audio 2.1 speaker diarization pipeline: principle, benchmark, and recipe},
  author    = {Hervé Bredin},
  year      = {2023},
  booktitle = {Interspeech 2023},
  pages     = {1983--1987},
  doi       = {10.21437/Interspeech.2023-105},
  issn      = {2958-1796},
}

@inproceedings{Sortformer,
  title={Sortformer: A Novel Approach for Permutation-Resolved Speaker Supervision in Speech-to-Text Systems},
  author={Tae Jin Park and Ivan Medennikov and Kunal Dhawan and Weiqing Wang and He Huang and Nithin Rao Koluguri and Krishna C. Puvvada and Jagadeesh Balam and Boris Ginsburg},
  year={2024},
  url={https://api.semanticscholar.org/CorpusID:280148777}
}

@inproceedings{t-sot,
  title     = {{Streaming Multi-Talker ASR with Token-Level Serialized Output Training}},
  author    = {Naoyuki Kanda and Jian Wu and Yu Wu and Xiong Xiao and Zhong Meng and Xiaofei Wang and Yashesh Gaur and Zhuo Chen and Jinyu Li and Takuya Yoshioka},
  year      = {{2022}},
  booktitle = {{Interspeech 2022}},
  pages     = {{3774--3778}},
  doi       = {{10.21437/Interspeech.2022-7}},
  issn      = {{2958-1796}},
}

@misc{whisper,
  doi = {10.48550/ARXIV.2212.04356},
  url = {https://arxiv.org/abs/2212.04356},
  author = {Radford, Alec and Kim, Jong Wook and Xu, Tao and Brockman, Greg and McLeavey, Christine and Sutskever, Ilya},
  title = {Robust Speech Recognition via Large-Scale Weak Supervision},
  publisher = {arXiv},
  year = {2022},
  copyright = {arXiv.org perpetual, non-exclusive license}
}

@INPROCEEDINGS{fastconformer,
  author={Dima Rekesh et al.},
  booktitle={2023 IEEE Automatic Speech Recognition and Understanding Workshop (ASRU)}, 
  title={Fast Conformer With Linearly Scalable Attention For Efficient Speech Recognition}, 
  year={2023},
  volume={},
  number={},
  pages={1-8},
  doi={10.1109/ASRU57964.2023.10389701}}

@misc{Silero,
  author = {Silero Team},
  title = {Silero VAD: pre-trained enterprise-grade Voice Activity Detector (VAD), Number Detector and Language Classifier},
  year = {2024},
  publisher = {GitHub},
  journal = {GitHub repository},
  howpublished = {\url{https://github.com/snakers4/silero-vad}},
  commit = {insert_some_commit_here},
  email = {hello@silero.ai}
}

@misc{canary-v2,
      title={Canary-1B-v2 \& Parakeet-TDT-0.6B-v3: Efficient and High-Performance Models for Multilingual ASR and AST}, 
      author={Monica Sekoyan and Nithin Rao Koluguri and Nune Tadevosyan and Piotr Zelasko and Travis Bartley and Nikolay Karpov and Jagadeesh Balam and Boris Ginsburg},
      year={2025},
      eprint={2509.14128},
      archivePrefix={arXiv},
      primaryClass={cs.CL},
      url={https://arxiv.org/abs/2509.14128}, 
}
\end{document}